\documentclass[prl,twocolumn,amsmath,amssymb,superscriptaddress, bibnotes]{revtex4-2}

\usepackage{graphicx}
\usepackage{dcolumn}
\usepackage{bm}
\usepackage{color}

\begin{document}

\title{Observation of antiferromagnetic order as odd-parity multipoles\\ inside the superconducting phase in CeRh$_{2}$As$_{2}$}

\author{Mayu~Kibune}

\affiliation{Department of Physics, Kyoto University, Kyoto 606-8502, Japan}

\author{Shunsaku~Kitagawa}

\email{kitagawa.shunsaku.8u@kyoto-u.ac.jp}
\affiliation{Department of Physics, Kyoto University, Kyoto 606-8502, Japan}

\author{Katsuki~Kinjo}
\author{Shiki~Ogata}
\author{Masahiro~Manago}
\email{Present address: Shimane University}
\author{Takanori~Taniguchi}
\email{Present address: Tohoku University}
\author{Kenji~Ishida}

\affiliation{Department of Physics, Kyoto University, Kyoto 606-8502, Japan}

\author{Manuel~Brando}
\author{Elena~Hassinger}
\author{Helge Rosner}
\author{Christoph~Geibel}
\author{Seunghyun~Khim}

\affiliation{Max Planck Institute for Chemical Physics of Solids, D-01187 Dresden, Germany}

\date{\today}

\begin{abstract}
Spatial inversion symmetry in crystal structures is closely related to the superconducting (SC) and magnetic properties of materials.
Recently, several theoretical proposals that predict various interesting phenomena caused by the breaking of the local inversion symmetry have been presented.
However, experimental validation has not yet progressed owing to the lack of model materials.
Here we present evidence for antiferromagnetic (AFM) order in CeRh$_{2}$As$_{2}$ (SC transition temperature $T_{\rm SC} \sim 0.37$~K), wherein the Ce site breaks the local inversion symmetry. 
The evidence is based on the observation of different extents of broadening of the nuclear quadrupole resonance spectrum at two crystallographically inequivalent As sites.
This AFM ordering breaks the inversion symmetry of this system, resulting in the activation of an odd-parity magnetic multipole.
Moreover, the onset of antiferromagnetism $T_{\rm N}$ within an SC phase, with $T_{\rm N} < T_{\rm SC}$, is quite unusual in systems wherein superconductivity coexists or competes with magnetism.
Our observations show that CeRh$_{2}$As$_{2}$ is a promising system to study how the absence of local inversion symmetry induces or influences unconventional magnetic and SC states, as well as their interaction.
\end{abstract}

\maketitle

In most superconductors, the inversion symmetry of crystal structure (global inversion symmetry) is preserved.
Hence, superconducting (SC) symmetry can be classified into even-parity spin-singlet state or odd-parity spin-triplet state.
However, such a parity classification no longer holds in superconductors without global inversion symmetry (no inversion center in the crystal structure); instead, parity-mixed SC states are realized\cite{V.M.Edelstein_JETP_1989}. 
In such a state, the spin-triplet component is induced by the spin-orbit interaction, and fascinating phenomena are predicted, such as existence of a considerably large upper critical magnetic field $H_{\rm c2}$ due to the absence of Pauli-limiting effect\cite{D.F.Agterberg_PhysicaC_2003,S.Fujimoto_JPSJ_2007}. 
In fact, a large $H_{\rm c2}$ for a magnetic field parallel to the inversion breaking direction ($c$ axis) was observed in heavy-fermion (HF) superconductors CePt$_3$Si and Ce$M$Si$_3$ ($M$ =Rh and Ir) without global inversion symmetry\cite{E.Bauer_PRL_2004,N.Kimura_PRL_2005,I.Sugitani_JPSJ_2006}.
In contrast, for superconductors with local inversion breaking but preserved global inversion symmetry (no inversion center at particular atoms), recent theoretical studies have revealed that the odd-parity SC states such as pair density wave state can be stabilized under magnetic fields even if only a pairing interaction in the spin-singlet channel exists\cite{T.Yoshida_PRB_2012}.

\begin{figure}[!tb]
\includegraphics[width=8.2cm,clip]{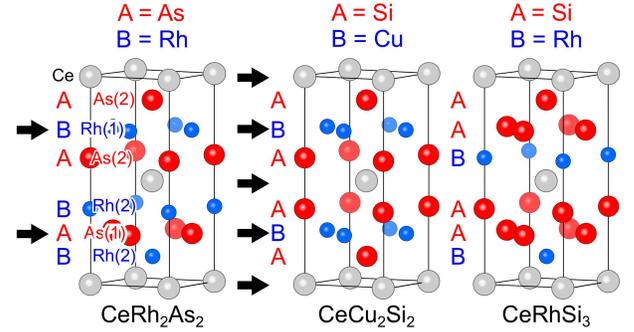}
\caption{Crystal structure of CeRh$_{2}$As$_{2}$(left), CeCu$_{2}$Si$_{2}$(center), and CeRhSi$_{3}$(right).
To clarify stack order, each element is assigned an A and a B and displayed on the left.
The arrows indicate the position of the inversion center.}
\label{Fig.1}
\end{figure}

In addition, if magnetic moments are aligned antiferromagnetically between sublattices at magnetic sites without local inversion symmetry, the system's magnetic ordering state can be regarded as an odd-parity multipole order, in which magnetic and electric degrees of freedom are entangled\cite{H.Watanabe_PRB_2017}.
In such a case, non-reciprocal conductivity and the magnetoelectric effect are expected and have been intensively studied from experimental and theoretical perspectives\cite{Y.Yanase_JPSJ_2014,H.Watanabe_PRB_2017,R.Wakatsuki_SciAdv_2017,Y.Shiomi_PRL_2019}.     

The spatial inversion symmetry of crystal structures seems to be also important for the relationship between SC and magnetic phases.
In many Ce-based HF and Fe-based superconductors with global inversion symmetry, superconductivity and antiferromagnetism coexist in a narrow region near the antiferromagnetic (AFM) quantum critical point, according to the phase diagram, as a function of tuning parameters\cite{C.Pfleiderer_RMP_2009,K.Ishida_JPSJ_2009}.
In addition, in most presently known systems wherein superconductivity coexists or competes with magnetism, coexistence has only be observed for $T_{\rm N} > T_{\rm SC}$, there is no observation of the onset of antiferromagnetism within an SC phase, even in those cases where antiferromagnetism reappears once the SC states are suppressed by a magnetic field\cite{T.Park_PNAS_2008,S.Sachdev_PhysicaC_2010}.
In contrast, in systems without global inversion symmetry, the coexistence region seems to be wider\cite{T.Yasuda_JPSJ_2004,R.Settai_JPSJ_2011}.
However, there is no report on the onset of antiferromagnetism within an SC phase even in this system\cite{R.Settai_JPSJ_2011}.
For systems without local inversion symmetry, no works have been reported in the literature regarding the relationship between these two phases, because no model material has been established yet. 

The unusual phenomena emerging from the absence of inversion symmetry rely on a strong spin-orbit interaction\cite{S.Fujimoto_JPSJ_2007}. 
Furthermore, strong electron correlations are required to explore the regime of unconventional superconductivity and magnetism. 
HF systems based on $f$-electrons, $e.g.$ Ce-based systems, offer all these ingredients and are therefore likely the best candidates for the observation and study of such phenomena\cite{T.Yasuda_JPSJ_2004,T.Yoshida_PRB_2012}.

CeRh$_{2}$As$_{2}$ is a recently discovered superconductor whose $T_{\rm SC}$ is approximately 0.3~K\cite{S.Khim_arXiv_2021}, which is lower than that of our samples due to differences in sample quality and experimental methods.
A broad maximum in resistivity at approximately 40~K and a large specific-heat jump at $T_{\rm SC}$ are typical features of HF superconductors\cite{S.Khim_arXiv_2021}.
The crystal structure is of the tetragonal CaBe$_{2}$Ge$_{2}$-type with space group $P4/nmm$ (No.129, $D_{4h}^7$)\cite{R.Madar_JLCM_1987}.
In CeRh$_{2}$As$_{2}$, there are two crystallographically inequivalent As and Rh sites; As(1)[Rh(1)] is tetrahedrally coordinated by Rh(2)[As(2)] as shown in Fig.~\ref{Fig.1}.
The crystal structure looks similar to the structure of the typical HF superconductor CeCu$_{2}$Si$_{2}$ (ThCr$_2$Si$_2$-type structure)\cite{F.Steglich_PRL_1979}; the global spatial inversion is preserved in both compounds.
However, the stacking order of the block layers is different.
A Ce layer in CeRh$_2$As$_2$ is located between two different block layers (Rh-As-Rh and As-Rh-As), and there is no inversion center at the Ce site.
In contrast, a Ce layer in CeCu$_2$Si$_2$ is sandwiched between identical block layers (Si-Cu-Si).
The comparison of these two compounds and Ce$M$Si$_{3}$ ($M$ = Rh and Ir), in which global inversion symmetry is broken with a Ce atom sandwiched between Si-Si-$M$ block layers, provide an ideal basis to study how removing inversion symmetry affects unconventional SC and magnetic states.
For example, in CeRh$_{2}$As$_{2}$, the SC upper critical field $\mu_0H_{\rm c2}$ along the $c$ axis is much higher than that perpendicular to the $c$-axis and with $\mu_0H_{{\rm c2}\parallel c} \sim$ 14~T far exceeds the Pauli-limiting field $\mu_0H_{\rm P}$ estimated from the formula $\mu_0H_{\rm P} \sim 1.84~T_{\rm SC} \sim 0.6$~T\cite{S.Khim_arXiv_2021}.
A similar enhancement of $\mu_0H_{{\rm c2}\parallel c}$ was observed in Ce$M$Si$_{3}$\cite{R.Settai_JPSJ_2008} but was not observed in CeCu$_2$Si$_2$\cite{S.Kittaka_PRB_2016}. 
Rather, $\mu_0H_{\rm c2}$ of CeCu$_2$Si$_2$ was understood well based on the Pauli-limiting scenario for $H \parallel a$ and $c$ axes\cite{S.Kitagawa_PRL_2018}.
In addition to the large $H_{\rm c2}$, in CeRh$_{2}$As$_{2}$, an SC--SC phase transition at approximately 4~T was reported for $H \parallel c$\cite{S.Khim_arXiv_2021}, which can be interpreted as a phase transition inside the SC phase from a low-field even parity to a high-field odd-parity state\cite{E.G.Schertenleib_arXiv_2021,K.Nogaki_arXiv_2021}. 
Such a phase transition is only expected in systems with locally broken inversion symmetry but has not been reported yet.

Apart from superconductivity, a further phase transition has been reported for CeRh$_{2}$As$_{2}$.
The specific heat shows a large anomaly at $T_{\rm SC}$, and a rather weak anomaly at $T_{0} \sim 0.4$~K.
$T_{0}$ shows a small decrease with increasing magnetic field applied to both directions but shifts to higher temperatures when magnetic fields perpendicular to the $c$ axis are higher than 5~T.
This peculiar field dependence of $T_{0}$ suggests that it is not of a simple magnetic order, but rather of a quadrupole-density-wave order\cite{D.Hafner_arXiv_2021}.

In this study, we report on nuclear quadrupole resonance (NQR) results at zero external field.
We observed broadening of the linewidth in NQR spectrum below 0.25~K, which is definitely lower than $T_{\rm SC} \sim$ 0.37~K.
From the site-dependent broadening of the NQR spectra, we conclude that this broadening indicates AFM order.
A magnetic transition inside an SC phase is a quite rare example, and we suggest that this anomalous coexistence might be related to local inversion symmetry breaking.

Single crystals of CeRh$_{2}$As$_{2}$ with a typical size of 3.0 $\times$2.0 $\times$0.75 mm$^3$ were grown using the Bi flux method\cite{S.Khim_arXiv_2021}.
$T_{\rm SC} = 0.37$~K was determined by the onset temperature of the SC diamagnetic signal from the ac susceptibility measurement using an NQR coil.
A conventional spin-echo technique was used for NQR measurements. 
Low-temperature NQR measurements below 1.5~K were performed using a $^{3}$He--$^{4}$He dilution refrigerator, in which the sample was immersed in the $^{3}$He--$^{4}$He mixture to avoid radio-frequency heating during measurements. 
The $^{75}$As-NQR spectra (nuclear spin $I~=~3/2$, nuclear gyromagnetic ratio $\gamma/2\pi~=~7.292$~MHz/T, and natural abundance 100\%) were obtained as a function of frequency at zero field.
The nuclear spin-lattice relaxation rate $1/T_1$ was determined by fitting the time variation of the nuclear magnetization probed with the spin-echo intensity after a saturation to a theoretical function for $I$ = 3/2\cite{A.Narath_PR_1967,D.E.MacLaughlin_PRB_1971}.
For $T_1$ measurements, we powdered single crystals and mixed them with Stycast 1266 to reduce the radio-frequency heating by the eddy current.
The $T_{\rm SC}$ of the powdered samples is almost the same as that of the single-crystalline samples.

\begin{figure*}[!t]
\includegraphics[width=17cm,clip]{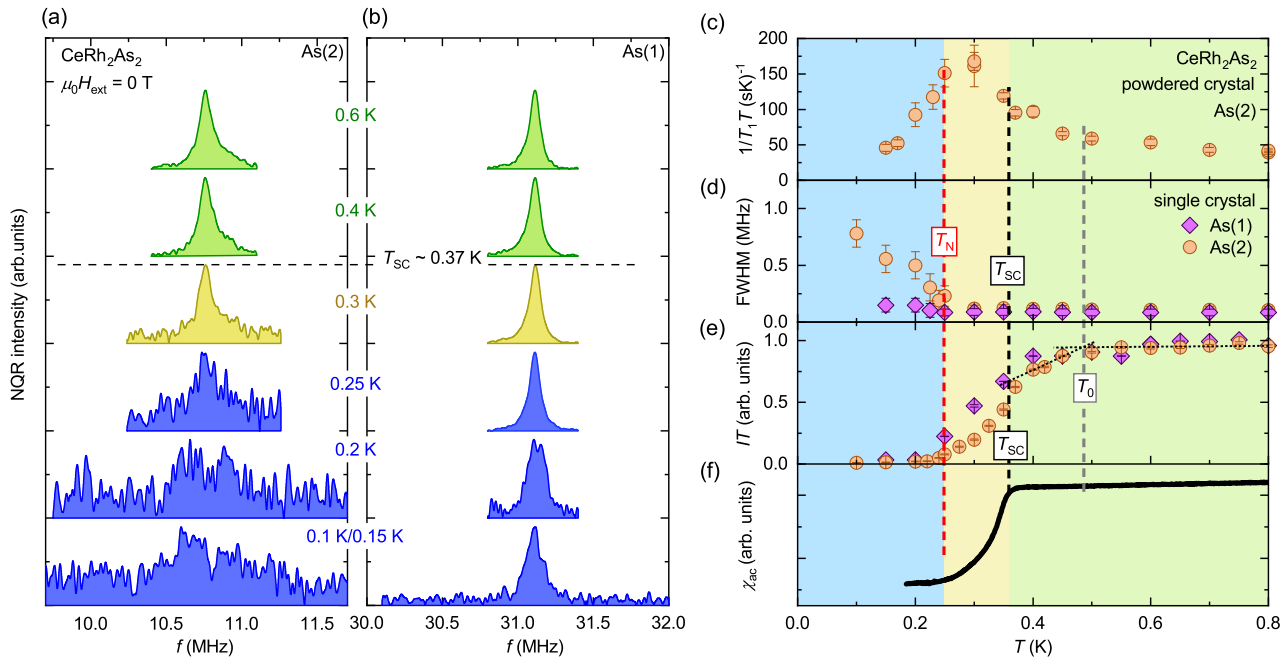}
\caption{Summary of the anomalies detected by NQR measurements.
$^{75}$As NQR spectrum of (a) As(2) and (b) As(1) sites at several temperatures in CeRh$_2$As$_2$.
Each spectrum is shifted to avoid overlapping.
Temperature dependencies of (c) $1/T_1T$, (d) FWHM of NQR spectrum, (e) NQR intensity multiplied by temperature $I(T)T$ of each As site, and (f) ac susceptibility.
For comparison, $I(T)T$ at each site is normalized by its value at 1.0 K.
The characteristic temperatures $T_{\rm N}$, $T_{\rm SC}$, and $T_{0}$ are indicated by the dashed lines.
The errors of the FWHM and $IT$ are determined from the resolution of NQR
signals.}
\label{Fig.2}
\end{figure*}

As shown in Figs.~\ref{Fig.2} (a) and~\ref{Fig.2} (b), we observed two NQR signals at $\sim$ 10.75 and $\sim$ 31.1~MHz owing to the presence of two As sites.
Even at zero magnetic field, the degeneracy of the nuclear energy levels is lifted owing to the electric quadrupole interaction with $I\ge$ 1.
The electric quadrupole Hamiltonian $\mathcal{H}_Q$ can be expressed as
\begin{equation}
\mathcal{H}_Q=\frac{h\nu_{\mathrm{Q}}}{6}\left[3I_{z}^{2}-I\left(I+1\right)+\frac{\eta}{2}\left(I_{+}^{2}+I_{-}^{2}\right)\right],
\label{eq.1}
\end{equation}
where $h \nu_{\mathrm{Q}}=\frac{3eQV_{zz}}{2I\left(2I-1\right)}$ and the asymmetry parameter $\eta = \left|\frac{V_{yy}-V_{xx}}{V_{zz}}\right|$.
In CeRh$_{2}$As$_{2}$, $\eta$ is zero at each As site because of the 4-fold symmetry of the atomic position; thus, it is difficult to assign each NQR signal to a specific site from the measurements.
We therefore performed a density functional theory (DFT)-based $ab~initio$ calculation of the electronic states, which provided considerably different NQR frequency $\nu_{\rm Q}$ values of 25.3 and 6.4~MHz for the As(1) and As(2), respectively.
A DFT calculation cannot take into account the strong correlation effects present in HF systems.
Therefore, the absolute values of $\nu_{\rm Q}$ are not reliable, but the ratio is expected to be representative.
Indeed the theoretically calculated $\nu_{\rm Q}$ ratio [As(1):As(2) $\sim$ 3.8:1] is not far from the experimental one (2.9:1).
Therefore, we ascribe the NQR signals at 31.1~MHz and 10.75~MHz to the As(1) and As(2) sites, respectively.

The sample quality of the present single-crystal CeRh$_2$As$_2$ is considered good compared to the single-crystal CeCu$_2$Si$_2$ we measured previously\cite{S.Kitagawa_PRB_2017,S.Kitagawa_PRL_2018}.
This is because the distribution of the NQR frequency $\nu_Q$ at the As(1) ($\Delta \nu_Q / \nu_Q \sim 2.7 \times 10^{-3}$) is smaller than that of the Cu-NQR signal in CeCu$_2$Si$_2$($\Delta \nu_Q / \nu_Q \sim 1.2 \times 10^{-2}$)\cite{S.Kitagawa_PRB_2017}; the local symmetry of the As(1) site in CeRh$_2$As$_2$ is similar to that of the Cu site in CeCu$_2$Si$_2$, as shown in Fig.~\ref{Fig.1}.
Here, $\Delta \nu_Q$ is estimated from the full width at the half maximum (FWHM) of the NQR signal.

We observed linewidth broadening of the NQR signal at unexpectedly low temperatures.
As shown in Fig.~\ref{Fig.2}, significant broadening of the linewidth at the As(2) site was observed below 0.25~K, while the change in the NQR spectrum at the As(1) site was marginal.
In contrast, the peak position in the two NQR spectra did not change significantly down to the lowest temperature (0.1~K).
To clarify the transition temperature, we compared the temperature dependencies of $1/T_1T$, the FWHM of the NQR spectrum, the NQR intensity multiplied by temperature $I(T)T$ of each As site and the ac susceptibility (right panel of Fig.~\ref{Fig.2}).
The ac susceptibility clearly decreased below $T_{\rm SC} = 0.37$~K owing to the SC diamagnetic signal.
In contrast, particularly at the As(2) site, $I(T)T$ started to decrease from $\sim$ 0.5~K, which is higher than $T_{\rm SC}$.
While the significant decrease in $I(T)T$ below $T_{\rm SC}$ is related to the SC diamagnetic effect for the RF pulses, the small decrease above $T_{\rm SC}$ is attributed to other aspects.
The most likely candidate is the phase transition denoted by $T_{0}$ observed in specific heat measurements\cite{S.Khim_arXiv_2021}.
The reduction in $IT$ at $T_0$ may be related to an increase in the nuclear spin-spin relaxation time $T_2$.
Because we only detected the decrease in the signal intensity, it is difficult to clarify the origin of this phase transition from the present work.
Thus, the most important result of our present NQR study is that, in addition to the anomalies at $T_{0} \sim 0.5$~K and $T_{\rm SC} =0.37$~K, there was an increase in the NQR linewidth below $T_{\rm N}$ = 0.25~K, which was very pronounced for the As(2) site but rather weak for the As(1) site.
Because the $T_1$ at both the As(1) and As(2) sites ($<$ 150 ms) was shorter than the repetition time of the radio-frequency-pulse ($>$ 1 s), the broadening is not due to the variation in $T_1$.
In addition, $1/T_1T$ also shows anomalies at these three transition temperatures; $1/T_1T$ began to increase at $T_0$, increased rapidly at $T_{\rm SC}$, and decreased below $T_{\rm N}$.
These results indicate that the linewidth broadening, SC transition, and $T_0$ anomaly are intrinsic properties of CeRh$_2$As$_2$.

The observed linewidth broadening below 0.25~K indicates the occurrence of internal magnetic field.
In general, linewidth broadening is caused by the appearance of small magnetic field or the increase in the distribution of electric field gradient (EFG).
Note that the NQR spectrum should be split when the magnetic field is sufficiently large.
One possible origin is the SC diamagnetic field owing to Meissner shielding.
However, NQR measurements were performed at zero field, and thus, the broadening of the NQR linewidth cannot be due to the SC diamagnetic field. 
Another possibility is due to an increase in the distribution of EFG.
The EFG at nuclear sites, related to $\nu_{\rm Q}$, is affected by changes in the charge distribution associated with a structural phase transition, charge order, charge density wave, or quadrupole order.
In such cases, the value of $\nu_{\rm Q}$ usually changes\cite{K.Arai_JPCM_2002,T.Koyama_JMMM_2004,M.Manago_JPSJ_2021}, but not in this case.
In addition, naively, the change in the distribution of $\nu_{\rm Q}$ should be proportional to the value of $\nu_{\rm Q}$ or to the distribution of $\nu_{\rm Q}$ in the normal (paramagnetic) state.
The observed site-dependent linewidth broadening $(\Delta \nu_{Q}^{\rm As(2)}/\Delta \nu_Q^{\rm As(1)} \sim 5.3 )$ was larger than either of $[\nu_Q^{\rm As(2)}/\nu_Q^{\rm As(1)} \sim 0.3$, and $(\Delta \nu_Q / \nu_Q)^{\rm As(2)}/(\Delta \nu_Q / \nu_Q)^{\rm As(1)} \sim 3.7$ in the normal state].
These results are also inconsistent with the scenario of increased linewidth due to the distribution of EFG.
Furthermore, if the atomic position of As(2) moves randomly below 0.25 K, Rh(2), which is in an equivalent Wyckoff position to As(2), would move in the same manner, and the linewidth of As(1) should be affected since As(1) is surrounded by Rh(2).
Therefore, the change in the distribution of EFG can be ruled out as the reason for the linewidth broadening.

\begin{figure*}[!tb]
\includegraphics[width=17cm,clip]{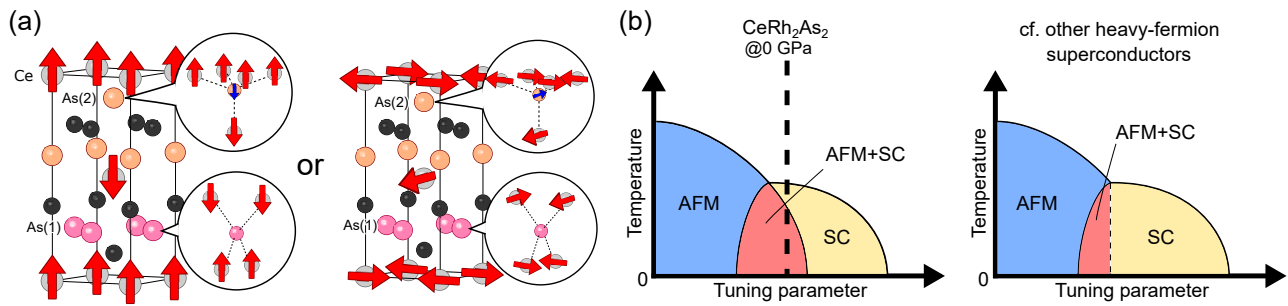}
\caption{Information on the magnetic phase.
(a) Possible magnetic structures below $T_{\rm N}$ and (b) schematic image of expected phase diagram for CeRh$_{2}$As$_{2}$.
The dashed line is the expected position of CeRh$_2$As$_2$ at ambient pressure.
For comparison, the schematic of the phase diagram observed in many HF superconductors is provided.
The red and blue arrows indicate the direction of the magnetic moment at the Ce site and that of the internal magnetic field at the As site, respectively.
AFM(SC) mean antiferromagnetic(superconducting) phase.}
\label{Fig.3}
\end{figure*}

Rather, the site-dependent linewidth broadening observed below 0.25~K is understandable in terms of an AFM order.
In addition to the electric quadrupole interaction, the Kramers degeneracy of the nuclear spin levels is lifted by a magnetic field due to the Zeeman interaction. 
Total effective Hamiltonian is the sum of two Hamiltonians and expressed as
\begin{align}
\mathcal{H} &= \mathcal{H}_z + \mathcal{H}_Q \nonumber \\
&= - \gamma \hbar (1 + K)I\cdot H + \mathcal{H}_Q,
\end{align}
where $K$ is the Knight shift, and $H$ is a magnetic field.
The increase in the linewidth of an NQR spectrum is roughly proportional to the internal magnetic field at each As site.
In our experiments, significant broadening was observed only at the As(2) site; the increase in the linewidth at the As(1) site was small.
Thus, an AFM order is the most plausible explanation since the internal magnetic field can be canceled out at the symmetric site due to the superposition of the transferred field of different Ce neighbors.
Although it is difficult to estimate the magnitude of the off-diagonal term of transferred hyperfine tensor which is related to an internal field causes by an AFM ordering, we want to know only whether the internal magnetic field originating from the Ce 4$f$ moments is cancelled out.
Therefore, we estimated the internal magnetic field at each As site using the classical dipole interaction to consider possible magnetic structures.
The dipolar magnetic field at each As site $\bm{H}_{\rm int}^{\rm As}$ from the Ce 4$f$ moments can be expressed as,
\begin{align}
\bm{H}_{\rm int}^{\rm As} &= \sum_i \bm{H}^{\rm dip}_i ,\\
\bm{H}^{\rm dip}_i(\bm{r}_i) &= -\nabla \frac{\bm{\mu}_i \cdot \bm{r_i}}{4\pi \bm{r}_i^{3}},
\end{align}
where $\bm{\mu}_i$ is a magnetic moment at $i$th Ce site, and $\bm{r}_i$ is the relative position vector of $i$th Ce site from the As site.
From these estimations, the A-type AFM (in-plane ferromagnetic and inter-plane antiferromagnetic) order with magnetic moments parallel to the $c$ axis or a helical order with in-plane moments as shown in Fig.~\ref{Fig.3} (a) are promising candidates for the magnetic structure.
In both magnetic structures, the internal magnetic field generated by the Ce 4$f$ moments is canceled out at the As(1) site, while a non-zero internal magnetic field appears at the As(2) site.
To determine the magnetic structure, diffraction experiments such as those of neutron diffraction are necessary.

These results indicate that the coexistence of superconductivity, $T_{0}$ anomaly, and AFM order is realized in CeRh$_{2}$As$_{2}$.
Such a multiple coexistence system is quite unusual.
In addition, we noted that the onset of the AFM order inside the SC phase is unique.
As mentioned above, in most systems where superconductivity coexists with AFM order, superconductivity occurs below $T_{\rm N}$, and the AFM order disappears when $T_{\rm SC}$ is higher than $T_{\rm N}$ [see the right panel of Fig.\ref{Fig.3} (b)].
This asymmetric phase diagram is explained by the difference in the position on the Fermi surface (FS), where the respective energy gap develops.
In metallic AFM systems, a folding gap opens only on the FSs connected by the magnetic wave vectors in the AFM state, and the remaining electrons can condense into an SC state. 
In contrast, the SC gaps open on almost all FSs in the SC state, which makes the formation of an AFM state of heavy fermions rather difficult.
To overcome this situation, localized magnetic moments or magnetic fields are required\cite{W.A.Fertig_PRL_1977,M.Kenzelmann_Science_2008,S.Raymond_JPSJ_2014,G.Prando_PRB_2013}.
In contrast to previous cases, in CeRh$_{2}$As$_{2}$, the AFM order in the SC phase is realized without introducing magnetic impurity.
We suggest that the presence of the $k$-dependent fictional field originating from the Rashba-type interaction\cite{S.Fujimoto_JPSJ_2007,T.Yoshida_PRB_2012} may stabilize the AFM order inside the SC state.
Because we only performed measurements at ambient pressure, it is important to clarify the overall phase diagram of CeRh$_2$As$_2$ via pressure experiments; however, we expect the phase diagram to be as shown in the left panel of Fig.~\ref{Fig.3} (b).

Finally, we comment on the magnetic structures suggested from the $^{75}$As-NQR spectra, which are shown in Fig.~\ref{Fig.3} (a).
Notably, neither AFM structures matches before and after the inversion operation at the As(1) or Rh(1) layer, because a magnetic moment is not reversed by the inversion operation.
This means that the AFM transition breaks the global inversion symmetry, resulting in odd-parity magnetic multipoles becoming active in the ground state.
It was predicted that an exotic superconductivity analogous to that in the Fulde--Ferrell--Larkin--Ovchinnikov state could be stabilized without applying external magnetic fields when such a multipole coexists with superconductivity\cite{S.Sumita_PRB_2016}. 
We point out the possibility that the unusual coexistence can be explained by this scenario.
Until now, the superconductivity coexisting with an odd-parity magnetic multipole has never been reported.
Therefore, CeRh$_2$As$_2$ provides a promising platform to study the relationship between superconductivity and unconventional multipoles.
   
In conclusion, we have performed $^{75}$As-NQR measurements on the recently discovered HF superconductor CeRh$_{2}$As$_{2}$ to investigate the electronic properties at low temperatures.
We observed broadening of the linewidth in the NQR spectra below 0.25~K.
From the site-dependent linewidth broadening, we deduced the onset of the AFM order and discussed possible AFM structures.
Because a clear SC transition was observed at 0.37~K, the AFM order was located inside the SC phase, with $T_{\rm N} < T_{\rm SC}$ being a very unusual situation for coexistence of antiferromagnetism and superconductivity.
This coexistence in CeRh$_{2}$As$_{2}$ might be related to its unique crystal structure with the locally broken inversion symmetry at the Ce site, which induces a fictitious Zeeman field. 
We point out that the SC state in CeRh$_{2}$As$_{2}$ might be a very unconventional one resulting from the interaction with an odd-parity multipole within a locally non-centrosymmetric crystal structure.
Our finding provides a new avenue for investigating new types of unconventional superconductivity.

\section*{acknowledgments}
The authors would like to thank Y. Maeno, S. Yonezawa, Y. Yanase for valuable discussions.
This work was partially supported by the Kyoto University LTM Center, and Grants-in-Aid for Scientific Research (KAKENHI) (Grants No. JP15H05745, No. JP17K14339, No. JP19K14657, No. JP19H04696, and No. JP20H00130). 
C. G. and E. H. acknowledge support from the DFG through Grant No. GE 602/4-1 Fermi-NEST.

M.K. and S.K. contributed equally to this work.

\end{document}